\documentstyle[12pt]{article}
\newcommand{\bd}{\begin{document}}
\newcommand{\ed}{\end{document}}
\newcommand{\bc}{\begin{center}}
\newcommand{\ec}{\end{center}}
\newcommand{\be}{\begin{eqnarray}}
\newcommand{\ee}{\end{eqnarray}}
\renewcommand{\thefootnote}{\alph{footnote}}
\newcommand{\se}{\section}
\newcommand{\sse}{\subsection}
\newcommand{\bi}{\bibitem}
\def\figcap{\section*{Figure Captions\markboth
     {FIGURECAPTIONS}{FIGURECAPTIONS}}\list
     {Figure \arabic{enumi}:\hfill}{\settowidth\labelwidth{Figure 999:}
     \leftmargin\labelwidth
     \advance\leftmargin\labelsep\usecounter{enumi}}}
\let\endfigcap\endlist \relax
\input psfig

\begin{document}

\begin{titlepage}

\vskip 0.5in
\null
\begin{center}
\vspace{.15in}
{\Large \bf
Branching ratios of $B_s\to (K^+K^-,K^0\bar{K}^0)$
and \\
$B_d\to \pi^+ \pi^-$ and
determination of
\boldmath{$\gamma (\phi_3)$}
}\\
\vspace{1.0cm}  \par
\vskip 2.1em
{\large
  \begin{tabular}[t]{c}
{\bf Chuan-Hung Chen$^a$ and C.~Q.~Geng$^b$}
\\
\\
   {\sl ${}^a$Department of Physics, National Cheng Kung University}
\\   {\sl  $\ $Tainan, Taiwan,  Republic of China }
\\
\\
{\sl ${}^b$Department of Physics, National Tsing Hua University}
\\  {\sl  $\ $ Hsinchu, Taiwan, Republic of China }
\\
   \end{tabular}}
\par \vskip 5.3em

\date{\today}
{\Large\bf Abstract}

\end{center}

We explored various cases for the branching ratios (BRs) of
$B_s\to K^+K^-$, $B_s\to K^0\bar{K}^0$ and $B_d\to \pi^+ \pi^-$
decays. We study the possibility of determining $\gamma $ by using
the following the measurements: (a) BRs of $B_s\to K^+K^-$,
and $B_s\to K^0\bar{K}^0$; (b) the ratio of
direct CP asymmetries in $B_d\to \pi^+ \pi^-$ and $B_s\to
K^+K^-$;
(c) the mix-induced CP asymmetry in $B_d\to \pi^+ \pi^-$; and (d) the
angle of $\beta$.

\end{titlepage}

One of the purposes in present and future B physics experiments is to
determine the CP violating angles of the Cabibbo-Kobayashi-Maskawa (CKM)
matrix \cite{CKM} induced from the three-generation quark mixings and
described by the three angles $\alpha $, $\beta $ and $\gamma $ or
$\phi_{2}$%
, $\phi_{1}$ and $\phi _{3}.$ With the unitary property, they
satisfy the triangular identity of $\alpha +\beta +\gamma =\pi$.
In the literature, the time-dependent rate asymmetry of the
$B_{d}\rightarrow \pi ^{+}\pi ^{-}$ decay is suggested to
determine the angle $\alpha $. However, large uncertainties from
the inevitable pollution of the penguin-topology make this
procedure be limited in experiments although some methods have
been suggested to rescue them \cite{Cure1, Cure2}. Nevertheless,
the similar approach applied to $B\rightarrow J/\Psi K_{s}$ for
extracting $\beta $ \cite {CS} is clean both theoretically and
experimentally. The angle $\gamma $
determination by the branching ratios (BRs) and direct CP asymmetries of $%
B\rightarrow K\pi $ decays has been also proposed \cite{KPI,KPI1}.

It is known that in the SU(3) flavor symmetry limit, there are relations
between different decay amplitudes \cite{GHLR}. In particular, one has
\begin{eqnarray}
{\cal A}(B_{s}\rightarrow K^{0}\bar{K}^{0}) &=&\frac{V_{t}}{\lambda _{t}}%
{\cal A}(B_{d}\rightarrow K^{0}\bar{K}^{0}),  \nonumber \\
{\cal A}(B_{s}\rightarrow K^{+}K^{-})+{\cal A}(B_{s}\rightarrow
K^{0}\bar{K}%
^{0}) &=&\frac{V_{u}}{\lambda _{u}}\left( {\cal A}\left( B_{d}\rightarrow
\pi ^{+}\pi ^{-}\right) +{\cal A}\left( B_{d}\rightarrow K^{0}\bar{K}%
^{0}\right) \right) .  \label{KK}
\end{eqnarray}
Furthermore, if the effects from annihilation contributions are
negligible, we have new relations
\begin{eqnarray}
{\cal A}(B_{s}\rightarrow K^{+}K^{-})&=&{\cal A}(B_{d}\rightarrow K^{+}\pi
^{-})\,,  \nonumber \\
{\cal A}(B_{s}\rightarrow K^0\bar{K}^0)&=&{\cal A}(B_u^-\to K^0\pi^-)\,.
\label{bkpi}
\end{eqnarray}
Especially, except the different CKM matrix elements, under the U-spin
transformation, since the subgroup of SU(3) describes the interchange of d-
and s-quark, the transition matrix elements associated with various
topologies in $B_{d}\rightarrow \pi ^{+}\pi ^{-}$ and $B_{s}\rightarrow
K^{+}K^{-}$ decays are related to each other. By using the connected
relations and combining the mix-induced and direct CP asymmetries in both $%
B_{d}\rightarrow \pi ^{+}\pi ^{-}$ and $B_{s}\rightarrow K^{+}K^{-}$ decays,
Fleischer in Ref. \cite{Fleischer2} has proposed strategies to determine $%
\gamma$.

In this paper, we will first evaluate the branching ratios (BRs) of $B_s\to
(K^+K^-$, $K^0\bar{K}^0)$ and $B_d\to \pi^+ \pi^-$ and then analyze the
possibility of extracting angle $\gamma $ through $BR(B_{s}\to KK)$ and the
relevant measurements in $B_{d}\rightarrow \pi ^{+}\pi ^{-}$. Other
approaches associated with $B_{s}\rightarrow KK$ can refer to \cite
{FD,DFN,Kim,FP}

We start by writing the amplitudes for $B_{s}\rightarrow K^+K^-$, $B_s\to
K^0%
\bar{K}^0$, $B_{d}\rightarrow \pi ^{+}\pi ^{-}$ and $B_{d}\rightarrow K^{0}%
\bar{K}^{0}$ decays generally as
\begin{eqnarray}
{\cal A(}B_{s}\rightarrow K^{+}K^{-}{\cal )} &=&V_{t}P_{t}+V_{c}P_{c}+V_{u}%
\left( P_{u}+T\right)  \nonumber \\
&=&V_{c}P_{ct}\left( 1+\left| \frac{V_{u}}{V_{c}}\right| Re^{i(\delta
+\gamma )}\right)  \label{apckck} \\
{\cal A(}B_{s}\rightarrow K^{0}\bar{K}^{0}{\cal )} &=&V_{c}P_{ct}\left( 1+%
\frac{V_{u}}{V_{c}}\frac{P_{ut}}{P_{ct}}\right)  \label{apnknk} \\
{\cal A(}B_{d}\rightarrow \pi ^{+}\pi ^{-}{\cal )} &=&\lambda
_{t}P_{t}^{\prime }+\lambda _{c}P_{c}^{\prime }+\lambda _{u}\left(
P_{u}^{\prime }+T^{\prime }\right)  \nonumber \\
&=&\lambda _{c}P_{ct}^{\prime }\left( 1-\left| \frac{\lambda _{u}}{\lambda
_{c}}\right| re^{i(\delta ^{\prime }+\gamma )}\right)  \label{appipi} \\
{\cal A(}B_{d}\rightarrow K^{0}\bar{K}^{0}{\cal )} &=&\lambda
_{c}P_{ct}^{\prime \prime }\left( 1+\frac{\lambda _{u}}{\lambda _{c}}\frac{%
P_{ut}^{\prime \prime }}{P_{ct}^{\prime \prime }}\right)  \label{apdnknk}
\end{eqnarray}
where $P_{q}^{(\prime ,\prime \prime )}$ and $T^{(\prime )}$ denote the
penguin- and tree-topology contributions, $P_{qq^{\prime }}^{(\prime ,\prime
\prime )}=P_{q}^{(\prime ,\prime \prime )}-P_{q^{\prime }}^{(\prime ,\prime
\prime )}$, $R(r)e^{i\delta ^{(\prime )}}=\left( P_{ut}^{(\prime
)}+T^{(\prime )}\right) /P_{ct}^{(\prime )}$, $\lambda
_{i}=V_{id}^{*}V_{ib}$%
, $V_{i}=V_{is}^{*}V_{ib}$ in which $V_{ij}$ are the CKM matrix elements and
they satisfy $\sum_{i}V_{i}(\lambda _{i})=0.$ With the Wolfenstein's
parametrization \cite{Wolfenstein}, we know that
\[
\begin{tabular}{lllll}
$\lambda _{t}=A\lambda ^{3}R_{t}e^{-i\beta },$ &  & $\lambda _{c}=-A\lambda
^{3},$ &  & $\lambda _{u}=A\lambda ^{3}R_{b}e^{-i\gamma },$ \\
$V_{t}=-A\lambda ^{2},$ &  & $V_{c}=A\lambda ^{2},$ &  & $V_{u}=A\lambda
^{4}R_{b}e^{-i\gamma }$%
\end{tabular}
\]
where
\[
\lambda \approx 0.22,\ A\approx 0.80,\ R_{b}\approx 0.36\,.
\]
In the SU(3) limit, one finds that
\begin{eqnarray}
P_{ct} &=&P_{ct}^{\prime }=P_{ct}^{\prime \prime }\,,  \nonumber \\
\frac{P_{ut}}{P_{ct}} &=&\frac{P_{ut}^{\prime \prime }}{P_{ct}^{\prime
\prime }}\,,  \nonumber \\
Re^{i\delta } &=&re^{i\delta ^{\prime }}.  \label{SU3}
\end{eqnarray}
However, by including SU(3)-breaking effects, the first relation in Eq.
(\ref
{SU3}) becomes
\begin{equation}
\frac{P_{ct}}{f_{K}F^{B_{s}K}(0)}\simeq \frac{P_{ct}^{\prime }}{f_{\pi
}F^{B_{d}\pi }(0)}\simeq \frac{P_{ct}^{\prime \prime }}{f_{K}F^{B_{d}K}(0)},
\label{su3-1}
\end{equation}
while the other two hold approximately, where we have set the light meson
masses to be zero, i.e., $M_{K}^{2}=0$ and $M_{\pi }^{2}=0$, and $f_{P}$ and
$F^{B_{q}P}(0)$ denote the $P$ meson decay constant and $B_{q}\rightarrow P$
decay form factor, respectively. Note that for parametrizing the SU(3)
broken effects in Eq. (\ref{su3-1}), we have used the concept of the
factorization assumption. In our following numerical analysis, we take $%
f_{\pi }=0.13$, $f_{K}=0.16$, $f_{B_{d}}=0.19$, $f_{B_{s}}=0.20$, $%
F^{B_{d}\pi }=0.3$ \cite{KLS}, $F^{B_{d}K}=0.35$ \cite{CLK}, and $%
F^{B_{s}K}=0.33\ GeV$ \cite{MS}. Since the last two relations in Eq. (\ref
{SU3}) are the ratios of the transition matrix elements, the nonperturbative
QCD effects can be reduced \cite{Fleischer2}. In order to estimate $\left|
P_{ut}\right| /\left| P_{ct}\right| $, we use $\delta =220^{0}$ , $\gamma
=76^{0}$, $r=8.0$ \cite{Fleischer2} and $Br\left( B_{d}\rightarrow \pi
^{+}\pi ^{-}\right) \simeq 5.43\times 10^{-6}$ \cite
{pipiBABAR,pipiBELLE,pipiCLEO}, then we get $\left| P_{ct}^{\prime }\right|
^{2}\approx 1.54\times 10^{-5}.$ Using the value of $\left| P_{ct}^{\prime
}\right| $ and Eqs. (\ref{apckck}) and (\ref{su3-1}), we obtain
\begin{eqnarray}
Br\left( B_{s}\rightarrow K^{+}K^{-}\right) \simeq 20.83\times 10^{-6}\,.
\end{eqnarray}
The result is consistent with that from Eq. (\ref{bkpi}), given by \cite
{GHLR,Fleischer2}
\begin{eqnarray}
Br\left( B_{s}\rightarrow K^{+}K^{-}\right) &\simeq &\frac{\tau _{B_{s}}}{%
\tau _{B_{d}}}\left( \frac{M_{B_{s}}}{M_{B_{d}}}\right) ^{3}\left( \frac{%
F^{B_{s}K}\left( 0\right) }{F^{B_{d}\pi }\left( 0\right) }\right)
^{2}Br\left( B_{d}\rightarrow K^{\pm }\pi ^{\mp }\right) ,  \nonumber \\
&\simeq &22.31\times 10^{-6}\,,  \label{bckck}
\end{eqnarray}
in which the measurement of $Br\left( B_{d}\rightarrow K^{\pm }\pi
^{\mp }\right) \simeq 18.2\times 10^{-6}$
\cite{pipiBABAR,pipiBELLE,pipiCLEO} is used$.$ From Eqs.
(\ref{apdnknk}) and (\ref{su3-1}) and the obtained value of
$\left| P_{ct}^{\prime }\right| $, the BR of $B_{d}\rightarrow
K^{0}\bar{K}^{0}$ is given by
\begin{eqnarray}
Br\left( B_{d}\rightarrow K^{0}\bar{K}^{0}\right) &=&\tau _{B_{d}}\frac{%
G_{F}^{2}M_{B_{d}}^{3}}{32\pi }\left| \lambda _{c}\right| ^{2}\left|
P_{ct}^{\prime \prime }\right| ^{2}\left( 1+R_{b}^{2}\left| \frac{%
P_{ut}^{\prime \prime }}{P_{ct}^{\prime \prime }}\right| ^{2}-2R_{b}\left|
\frac{P_{ut}^{\prime \prime }}{P_{ct}^{\prime \prime }}\right| {\cos}\gamma
{%
\cos}\theta \right) ,  \nonumber \\
&=&1.11\times 10^{-6}\left( 1+R_{b}^{2}\left| \frac{P_{ut}^{\prime \prime
}}{%
P_{ct}^{\prime \prime }}\right| ^{2}-2R_{b}\left| \frac{P_{ut}^{\prime
\prime }}{P_{ct}^{\prime \prime }}\right| \cos\gamma \cos\theta \right)
\end{eqnarray}
where the $\theta $ angle stands for the strong phase of $P_{ut}^{\prime
\prime }/P_{ct}^{\prime \prime }$. It is clear that the upper bound on $%
\left| P_{ut}^{\prime \prime }/P_{ct}^{\prime \prime }\right| $ occurs to
$\cos\gamma \cos\theta >0$. By taking $\theta =0^{0}$ and the limit of $%
Br\left( B_{d}\rightarrow K^{0}\bar{K}^{0}\right) \approx Br\left( B^{\pm
}\rightarrow K^{\pm }K^{0}\right) <2.5\times 10^{-6}$ \cite{GHLR,pipiBABAR},
we get $\left| P_{ut}^{\prime \prime }/P_{ct}^{\prime \prime }\right| <4$
and $\left| P_{ut}^{\prime \prime }/P_{ct}^{\prime \prime }\right| =1$ if
excluding rescattering effects. Hence, the second term in Eq. (\ref{apnknk})
can be neglected since $\left| V_{u}\right| /\left| V_{c}\right| \left|
P_{ut}^{\prime \prime }/P_{ct}^{\prime \prime }\right| \simeq \left|
V_{u}\right| /\left| V_{c}\right| \left| P_{ut}/P_{ct}\right| <0.069$, and
Eq. (\ref{apckck}) can be rewritten as
\begin{eqnarray}
{\cal A(}B_{s}\rightarrow K^{+}K^{-}{\cal )}={\cal A(}B_{s}\rightarrow
K^{0}%
\bar{K}^{0}{\cal )}\left( 1+\left| \frac{V_{u}}{V_{c}}\right| Re^{i(\delta
+\gamma )}\right) \,.
\end{eqnarray}
The CP averaged BR is given by
\begin{eqnarray}
\bar{B}r\left( B_{s}\rightarrow K^{+}K^{-}\right) &\equiv&\frac{Br\left(
B_{s}\rightarrow K^{+}K^{-}\right) +Br\left( \bar{B}_{s}\rightarrow
K^{-}K^{+}\right) }{2}  \nonumber \\
&=&Br\left( B_{s}\rightarrow K^{0}\bar{K}^{0}\right) \left( 1+\left| \frac{%
V_{u}}{V_{c}}\right| ^{2}R^{2}+2\left| \frac{V_{u}}{V_{c}}\right| R\cos%
\gamma \cos\delta \right) .  \label{brckck}
\end{eqnarray}

\begin{figure}[tbp]
\hspace{3.5cm} \psfig{figure=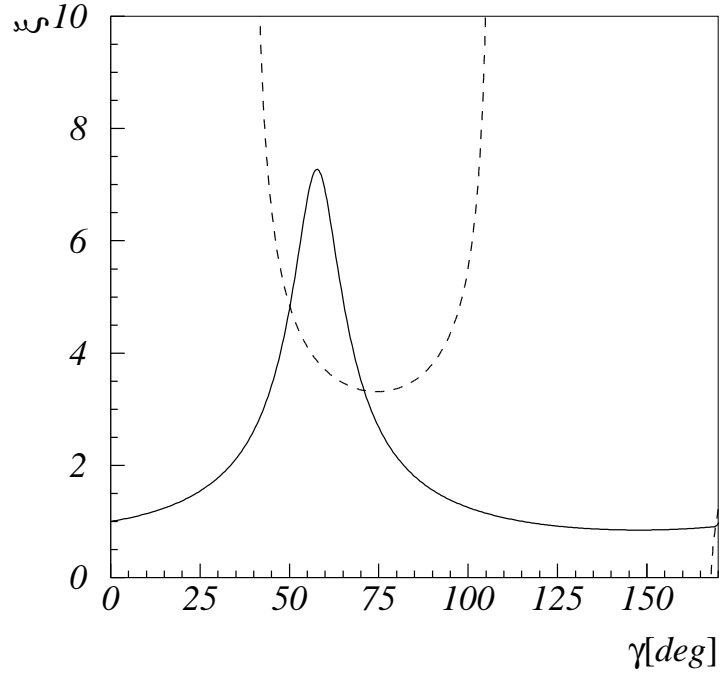,height=3.5in }
\caption{$\xi$ as function of $\gamma$ with the values of $%
A_{CP}^{mix}(B_d\rightarrow \pi^+ \pi^- )=0.45$, $A_{CP}^{dir}(B_d%
\rightarrow \pi^+ \pi^- )=-23\%$, $R_{CP}=1.4$ and $sin2\beta=0.60$. The
solid and dashed lines correspond to Eqs. (\ref{r1}) and (\ref{r2}),
respectively.}
\label{f1}
\end{figure}

 From Eqs. (\ref{apckck}$-$\ref{appipi}) and (\ref{su3-1}), we now
know that
there are four unknown parameters, $P_{ct}$, $r$, $\delta ^{\prime }$ and $%
\gamma $ in $B_{s}\rightarrow KK$ and $B_{d}\rightarrow \pi
^{+}\pi ^{-}$ decays. In terms of the analysis early, we see that
$\left| P_{ct}\right| $
can be fixed by the measurement of $Br\left( B_{s}\rightarrow K^{0}\bar{K}%
^{0}\right) $. As shown in Ref. \cite{Fleischer2}, the strong phase of $%
\delta ^{\prime }$ can be expressed as a function of $r$ and $\gamma $ by
using the mix-induced CP asymmetry from the time-dependent decaying rate
difference \cite{Fleischer1}, defined by $a_{CP}(t)\equiv \left( \Gamma
(B_{d}\rightarrow \pi ^{+}\pi ^{-})-\Gamma (B_{d}\rightarrow \pi ^{+}\pi
^{-})\right) /2,$ and explicitly, one has
\begin{eqnarray}
A_{CP}^{mix}\left( B_{d}\rightarrow \pi ^{+}\pi ^{-}\right) &=&{Im}\left(
e^{-i\phi _{d}}\frac{\bar{A}\left( B_{d}\rightarrow \pi ^{+}\pi ^{-}\right)
}{A\left( B_{d}\rightarrow \pi ^{+}\pi ^{-}\right) }\right) ,  \nonumber \\
&=&\frac{{sin}\phi _{d}-2\xi {\cos}\delta ^{\prime }{sin}\left( \phi
_{d}+\gamma \right) +\xi ^{2}{sin}\left( \phi _{d}+2\gamma \right) }{1-2\xi
{%
\cos}\delta ^{\prime }{\cos}\gamma +\xi ^{2}}.  \label{mixcp}
\end{eqnarray}
From the above equation, we easily obtain
\begin{equation}
2\xi cos\delta ^{\prime }=\xi ^{2}\rho +\omega  \label{delta}
\end{equation}
with
\begin{eqnarray*}
\rho &=&\frac{A_{CP}^{mix}\left( B_{d}\rightarrow \pi ^{+}\pi ^{-}\right)
-\sin \left( \phi _{d}+2\gamma \right) }{A_{CP}^{mix}\left( B_{d}\rightarrow
\pi ^{+}\pi ^{-}\right) {\cos}\gamma -\sin \left( \phi _{d}+\gamma \right)
},
\\
\omega &=&\frac{A_{CP}^{mix}\left( B_{d}\rightarrow \pi ^{+}\pi ^{-}\right)
-\sin \phi _{d}}{A_{CP}^{mix}\left( B_{d}\rightarrow \pi ^{+}\pi ^{-}\right)
{\cos}\gamma -\sin \left( \phi _{d}+\gamma \right) }
\end{eqnarray*}
where $\xi =\left| \lambda _{u}/\lambda _{c}\right| r$ and $\phi _{d}=2\beta
$ comes from the $B_{d}-\bar{B}_{d}$ mixing and its present status in
various experiments is listed as follows \cite{betaBELLE,betaBABAR,CDF}:
\begin{eqnarray}
\begin{tabular}{llllll}
${sin}\phi _{d}$ & $=$ & $0.58_{-0.34-0.10}^{+0.32+0.09}$ &  & $\left( {\rm
Belle}\right) ,$ &  \\
& $=$ & $0.59\pm 0.14\pm 0.05$ &  & $\left( {\rm BABAR}\right) ,$ &  \\
& $=$ & $0.79_{-0.44}^{+0.41}$ &  & $\left( {\rm CDF}\right) .$ &
\end{tabular}
\end{eqnarray}
In order to find the relationship between $r$ and $\gamma $ and
fix them, one can use the direct CP asymmetry in $B_{q}\rightarrow
PP$, defined by
\begin{equation}
A_{CP}^{dir}\left( B_{q}\rightarrow PP\right) =\frac{\Gamma \left(
B_{q}\rightarrow PP\right) -\Gamma \left( \bar{B}_{q}\rightarrow PP\right)
}{%
\Gamma \left( B_{q}\rightarrow PP\right) +\Gamma \left( \bar{B}%
_{q}\rightarrow PP\right) }\,,  \nonumber
\end{equation}
where $B_{q}$ can be $B_{d}$ or $B_{s}$ while $P$ is $\pi $ or $K$. From Eq.
(\ref{delta}), $A_{CP}^{dir}\left( B_{d}\rightarrow \pi ^{+}\pi ^{-}\right)
$
and the ratio of $A_{CP}^{dir}\left( B_{d}\rightarrow \pi ^{+}\pi
^{-}\right) $ to $A_{CP}^{dir}\left( B_{s}\rightarrow K^{+}K^{-}\right) $
\cite{Fleischer2}, one has
\begin{equation}
\xi =\sqrt{\frac{1}{h}\left( l\pm \sqrt{l^{2}-hk}\right) }\,,  \label{r1}
\end{equation}
and
\begin{equation}
\xi =\sqrt{\frac{1-tR_{CP}+t(1+R_{CP})\omega \cos\gamma }{%
t(R_{CP}-t)-t(1+R_{CP})\rho \cos\gamma }}\,,  \label{r2}
\end{equation}
respectively, where
\begin{eqnarray*}
\xi &=&\left| \lambda _{u}/\lambda _{c}\right| r \\
h &=&\rho ^{2}+C\left( 1-\rho {cos}\gamma \right) ^{2}, \\
k &=&\omega ^{2}+C\left( 1-\omega {cos}\gamma \right) ^{2}, \\
l &=&2-\rho \omega -C\left( 1-\rho {cos}\gamma \right) \left( 1-\omega
{cos}%
\gamma \right) , \\
C &=&\left( \frac{A_{CP}^{dir}\left( B_{d}\rightarrow \pi ^{+}\pi
^{-}\right) }{{sin}\gamma }\right) ^{2}, \\
R_{CP} &=&-\frac{A_{CP}^{dir}\left( B_{d}\rightarrow \pi ^{+}\pi ^{-}\right)
}{A_{CP}^{dir}\left( B_{s}\rightarrow K^{+}K^{-}\right) }, \\
t &=&\left| \frac{V_{u}}{V_{c}}\right| \left| \frac{\lambda _{c}}{\lambda
_{u}}\right|
\end{eqnarray*}
In Figure \ref{f1}, we show $\xi $ as a function of $\gamma $ in terms of
Eqs. (\ref{r1}) and (\ref{r2}) with $A_{CP}^{mix}\left( B_{d}\rightarrow \pi
^{+}\pi ^{-}\right) =0.45$, $A_{CP}^{dir}\left( B_{d}\rightarrow \pi ^{+}\pi
^{-}\right) =-0.23$ , $R_{CP}=1.4$ and sin2$\beta =0.60.$ From the figure,
we see that the crossing points between Eqs. (\ref{r1}) and (\ref{r2}) are
not unique. That is, it cannot completely settle down the $r$ and $\gamma $
with only the measurements of $A_{CP}^{mix}\left( B_{d}\rightarrow \pi
^{+}\pi ^{-}\right) $, $A_{CP}^{dir}\left( B_{d}\rightarrow \pi ^{+}\pi
^{-}\right) ,$ $R_{CP}$ and $\beta $. Therefore, one has to find another
independent relation to fix them.

To do this, we use both BRs of $B_{s}\rightarrow K^{+}K^{-}$ and $%
B_{s}\rightarrow K^{0}\bar{K}^{0}$ decays instead of using the
time-dependent CP asymmetry for the $B_{s}\rightarrow K^{+}K^{-}$ decay of
the approach in Ref. \cite{Fleischer2}. From Eqs. (\ref{brckck}) and (\ref
{delta}), we have
\begin{equation}
\xi =\sqrt{\frac{\left| \frac{\lambda _{u}}{\lambda c}\right| ^{2}\left(
R_{B}-1\right) -\left| \frac{V_{u}}{V_{c}}\right| \left| \frac{\lambda
_{u}}{%
\lambda c}\right| \omega {cos}\gamma }{\left| \frac{V_{u}}{V_{c}}\right|
^{2}+\left| \frac{V_{u}}{V_{c}}\right| \left| \frac{\lambda _{u}}{\lambda
_{c}}\right| \rho {cos}\gamma }}  \label{r3}
\end{equation}
where
\[
R_{B}=\frac{\bar{B}r\left( B_{s}\rightarrow K^{+}K^{-}\right) }{Br\left(
B_{s}\rightarrow K^{0}\bar{K}^{0}\right) }.
\]
In Eq. (\ref{bckck}), we have obtained the BR of $B_{s}\rightarrow
K^{+}K^{-}
$ in a model independent way. Similarly, we have
% According to Eq. (\ref{kk}) and
%following the similar approach, the BR for $B_{s}\rightarrow
%K^{0}\bar{K}^{0}$ can also be estimated by
\begin{eqnarray}
Br\left( B_{s}\rightarrow K^{0}\bar{K}^{0}\right)  &\simeq &\frac{\tau
_{B_{s}}}{\tau _{B_{d}}}\left( \frac{M_{B_{s}}}{M_{B_{d}}}\right) ^{3}\left|
\frac{V_{t}}{\lambda _{t}}\frac{F^{B_{s}K}(0)}{F^{B_{d}K}(0)}\right|
^{2}Br\left( B_{d}\rightarrow K^{0}\bar{K}^{0}\right)   \nonumber \\
&\simeq &24.65\times 10^{-6}\,,  \label{MI1}
\end{eqnarray}
where we have taken $Br\left( B_{d}\rightarrow K^{0}\bar{K}^{0}\right)
\simeq 1.33\times 10^{-6}$ \cite{CL}. We note that the decay rate of $%
B_{s}\to K^{0}\bar{K}^{0}$ can also be estimated based on the $SU(3)$
symmetry with the breaking effect and neglecting the small annihilation
contribution. Explicitly, we have
\begin{eqnarray}
Br(B_{s}\to K^{0}\bar{K}^{0}) &\simeq &\frac{\tau _{B_{s}}}{\tau _{B_{u}}}%
\left( \frac{M_{B_{s}}}{M_{B_{u}}}\right) ^{3}\left| \frac{F^{B_{s}K}(0)}{%
F^{B_{u}\pi }(0)}\right| ^{2}Br\left( B_{u}^{\pm }\rightarrow \pi ^{\pm
}K^{0}\right)   \nonumber \\
&\simeq &24.19\times 10^{-6}\,,
\end{eqnarray}
which agrees well with Eq. (\ref{MI1}), where we have used $Br\left(
B_{u}^{-}\rightarrow \pi ^{-}K^{0}\right) \simeq 21.0\times 10^{-6}$ \cite
{pipiBABAR,pipiBELLE,pipiCLEO}. From the estimations, we find that $Br\left(
B_{s}\rightarrow K^{+}K^{-}\right) $ prefers to being less than $Br\left(
B_{s}\rightarrow K^{0}\bar{K}^{0}\right) $. %Hence, we use the values of $%
%R_{B}\approx 0.91$ as the input value. The results with various choosing
%values are shown in Figures \ref{f2}-\ref{f4} and are summarized as
%follows:
It is clear that with specific values of relevant physical
observables, Eqs. (\ref{r1}), (\ref{r2}) and (\ref{r3}) can fix
$\gamma $. To illustrate our results, in Figure 2, we plot Eq.
(\ref{r3}) in the $\xi -\gamma $ plane with $R_{B}\simeq 0.91$ as
an input value and we find that $\gamma $ is about $73^{0}$. From
Eq. (\ref{brckck}), we see that the sign associated with
$\cos\delta \cos\gamma $ is positive. Since $\left| V_{u}\right|
^{2}R^{2}/\left| V_{c}\right| ^{2}\sim \lambda
^{4}R_{b}^{2}R^{2}$ which is in a few percent level and negligible, once $%
R_{B}<1$ $\left( >1\right) $ is measured, one can conclude that
$\cos\delta\cos\gamma <0$ $\left( >0\right) $. Moreover, through
the $\cos\delta$ described by Eq. (\ref{delta}), one can also
obtain the information
whether $\gamma $ is larger or less than $90^{0}$. In particular, for $%
R_{B}\simeq 1$ while $\cos \delta \cos \gamma \simeq 0$, one gets
$\gamma \simeq 90^{0}$ from Eq. (\ref{delta}). By fixing the
relevant observables except $R_{B}$, in Figures \ref{f3} and
\ref{f4}, we show how the angle of $\gamma $ is sensitive to the
value of $R_{B}$.
%%%%%%%%%%%%%%%%%%%%%%%%%%%%%%%%

It is worth to mention that the interference term in Eq. (\ref{appipi}),
associated with {\rm cos}$\delta ^{\prime }${\rm cos}$\gamma $ for $Br\left(
B_{d}\rightarrow \pi ^{+}\pi ^{-}\right) $, is negative. Therefore, in
contrast to the situation in the decay of $B_{s}\rightarrow K^{+}K^{-}$, the
BR of $B_{d}\rightarrow \pi ^{+}\pi ^{-}$ will have larger (smaller) values
if $\cos \delta ^{\prime }\cos \gamma <(>)0$. From Eq. (\ref{appipi}) one
can evaluate the BR of $B_{d}\rightarrow \pi ^{+}\pi ^{-}$
model-independently and the results are shown in Table \ref{brbpipi}.

\begin{table}[h]
\caption{The BR (in units of $10^{-6}$) of $B\rightarrow \pi ^{+}\pi ^{-}$
with $R(r)=8.0$, $\delta ^{({\prime })}=220^{0}$ and (I) $\gamma =70^{0}$,
(II) $\gamma =90^{0}$, and (III) $\gamma =110^{0}$. }
\label{brbpipi}
\begin{center}
\begin{tabular}{|c|c|c|}
\hline
$BR$ & {\rm model-independent} & {\rm experiment} \\ \hline
{\rm I} & $5.66$ & $4.1\pm 1.0\pm 0.7\cite{pipiBABAR}$ \\ \cline{1-2}
{\rm II} & $4.87$ & $5.9_{-2.1}^{+2.4}\pm 0.5\cite{pipiBELLE}$ \\
\cline{1-2}
{\rm III} & $4.08$ & $4.3_{-1.5}^{+1.6}\pm 0.5\cite{pipiCLEO}$ \\ \hline
\end{tabular}
\end{center}
\end{table}
\noindent Unfortunately, since the current accuracy of experiments
is limited, at the moment, we still cannot determine whether
$\cos\delta^{\prime}\cos\gamma $ is positive or negative but it
can be done in future B facilities. For a comparison, we display
the BRs of $B_{s}\rightarrow K^{+}K^{-}$ and $B_{d}\rightarrow \pi
^{+}\pi ^{-}$ as a function of $\gamma $ in Figure \ref{brpipi}.
 From the figure, it is interesting to find out that one of the
distributions increases with respect to the angle $\gamma $, while
the other one decreases.

In summary, we have studied the possibility of determining $\gamma $ by
using the virtual measurements of $A_{CP}^{mix}\left( B_{d}\rightarrow \pi
^{+}\pi ^{-}\right) $, $A_{CP}^{dir}\left( B_{d}\rightarrow \pi ^{+}\pi
^{-}\right) $, $\beta $, $R_{CP}$ and $R_{B}$. The first three physical
quantities can be measured precisely in the $e^{+}e^{-}$ machines at the $%
\Upsilon\left( 4S\right) $ resonance as well as hadronic ones such
as the Tevatron Run II. However, $R_{CP}$ and $R_{B}$ can be only
observed in the hadronic machines. It is known that the Tevatron
Run II has started a new physics run at $\sqrt{s}=2$ ${\rm TeV}$
and will collect a data sample of $2$ {\rm fb}$^{-1}$ in the first
two years \cite{CDF1}. At its initial phase with 10K of
$B_{s}\rightarrow KK$ events and by the relevant measurements in
$B_{d}\rightarrow \pi ^{+}\pi ^{-}$ decays, the observed $R_{B}$
could provide us a good opportunity to determine $\gamma$.

\newpage \noindent {\bf Acknowledgments}

This work was supported in part by the National Science Council of the
Republic of China under Contract Nos. NSC-89-2112-M-007-054 and
NSC-89-2112-M-006-033 and the National Center for Theoretical Science.

\newpage
\begin{figcap}
\item
$\xi$ as function of $\gamma$ with the values of $%
A_{CP}^{mix}(B_d\rightarrow \pi^+ \pi^- )=0.45$, $A_{CP}^{dir}(B_d%
\rightarrow \pi^+ \pi^- )=-23\%$, $R_{CP}=1.4$ and
$sin2\beta=0.60$. The solid and dashed lines correspond to Eqs.
(\ref{r1}) and (\ref{r2}), respectively.

\item
Same as Figure \ref{f1} but including Eq. (\ref{r3}) (dash-dotted
line) with $R_{B}=0.91$.

\item
$\xi$ as a function of $\gamma$ with  $%
A_{CP}^{mix}(B_d\rightarrow \pi^+ \pi^- )=0.1$, $A_{CP}^{dir}(B_d%
\rightarrow \pi^+ \pi^- )=-28\%$, $R_{CP}=1.8$, $sin2\beta=0.45$ and $%
R_{B}=1.02$. The curves are labeled as in Figure 2.

\item
Same as Figure \ref{f3} but with $R_{B}=1.18$.

\item
BRs (in units of $10^{-6}$) of $B_{s}\rightarrow K^+ K^-$ (dashed
line) and $B_{d}\rightarrow \pi^+ \pi^-$ (solid line) as a function of $%
\gamma$ with R(r)=8.0 and (a) $\delta=220^0$ and (b)
$\delta=40^0$.

\end{figcap}

\newpage

\begin{figure}[tbp]
\hspace{3.5cm} \psfig{figure=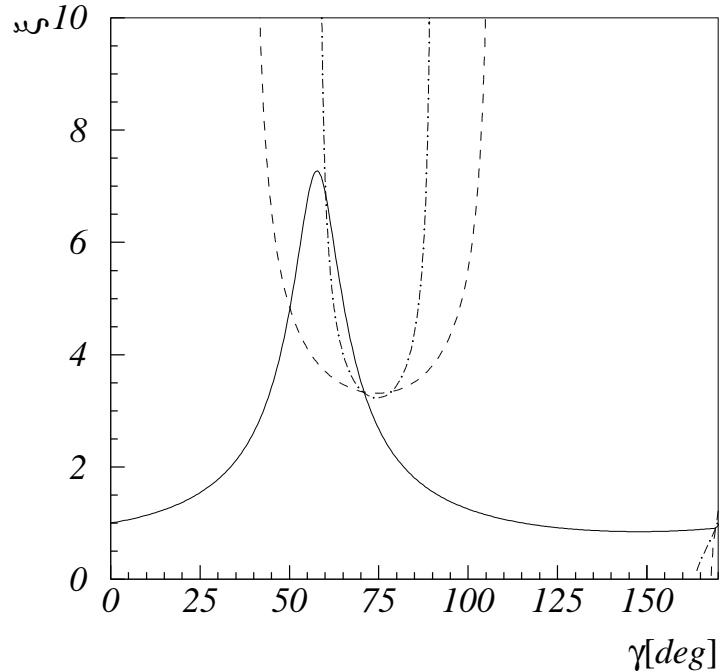,height=3.5in }
\caption{ Same as Figure \ref{f1} but including Eq. (\ref{r3}) (dash-dotted
line) with $R_{B}=0.91$. }
\label{f2}
\end{figure}

\begin{figure}[tbp]
\hspace{3.5cm} \psfig{figure=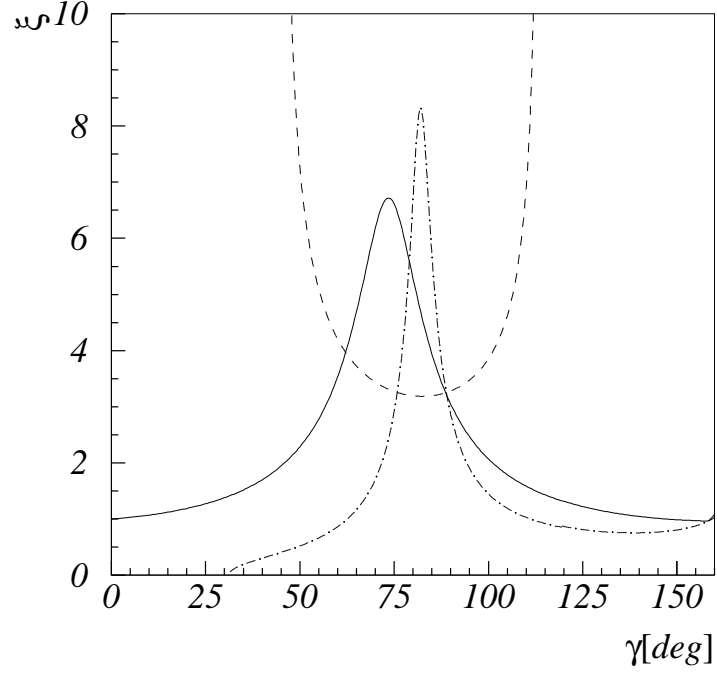,height=3.5in }
\caption{$\xi$ as a function of $\gamma$ with $A_{CP}^{mix}(B_d\rightarrow
\pi^+ \pi^- )=0.1$, $A_{CP}^{dir}(B_d\rightarrow \pi^+ \pi^- )=-28\%$, $%
R_{CP}=1.8$, $sin2\beta=0.45$ and $R_{B}=1.02$. The curves are labeled as in
Figure 2. }
\label{f3}
\end{figure}

\begin{figure}[tbp]
\hspace{3.5cm} \psfig{figure=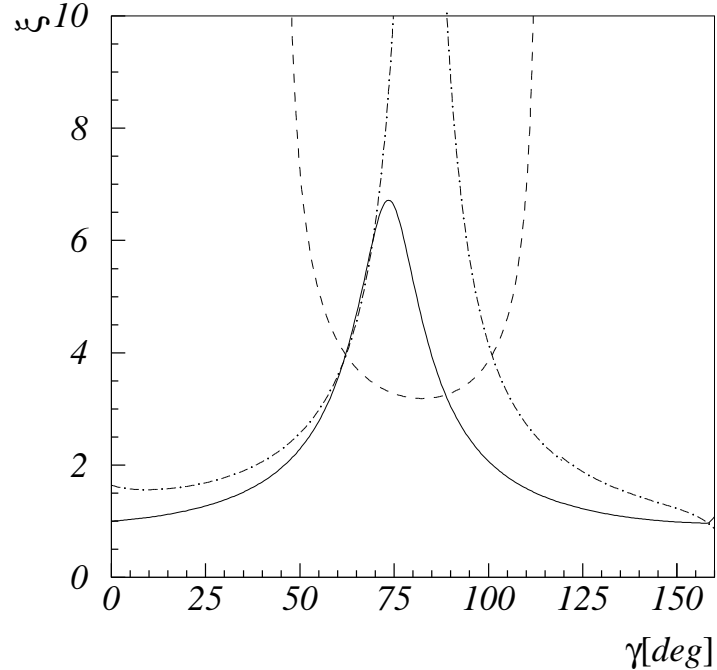,height=3.5in } \caption{Same
as Figure \ref{f3} but with $R_{B}=1.18$. } \label{f4}
\end{figure}

\begin{figure}[tbp]
\hspace{3.5cm} \psfig{figure=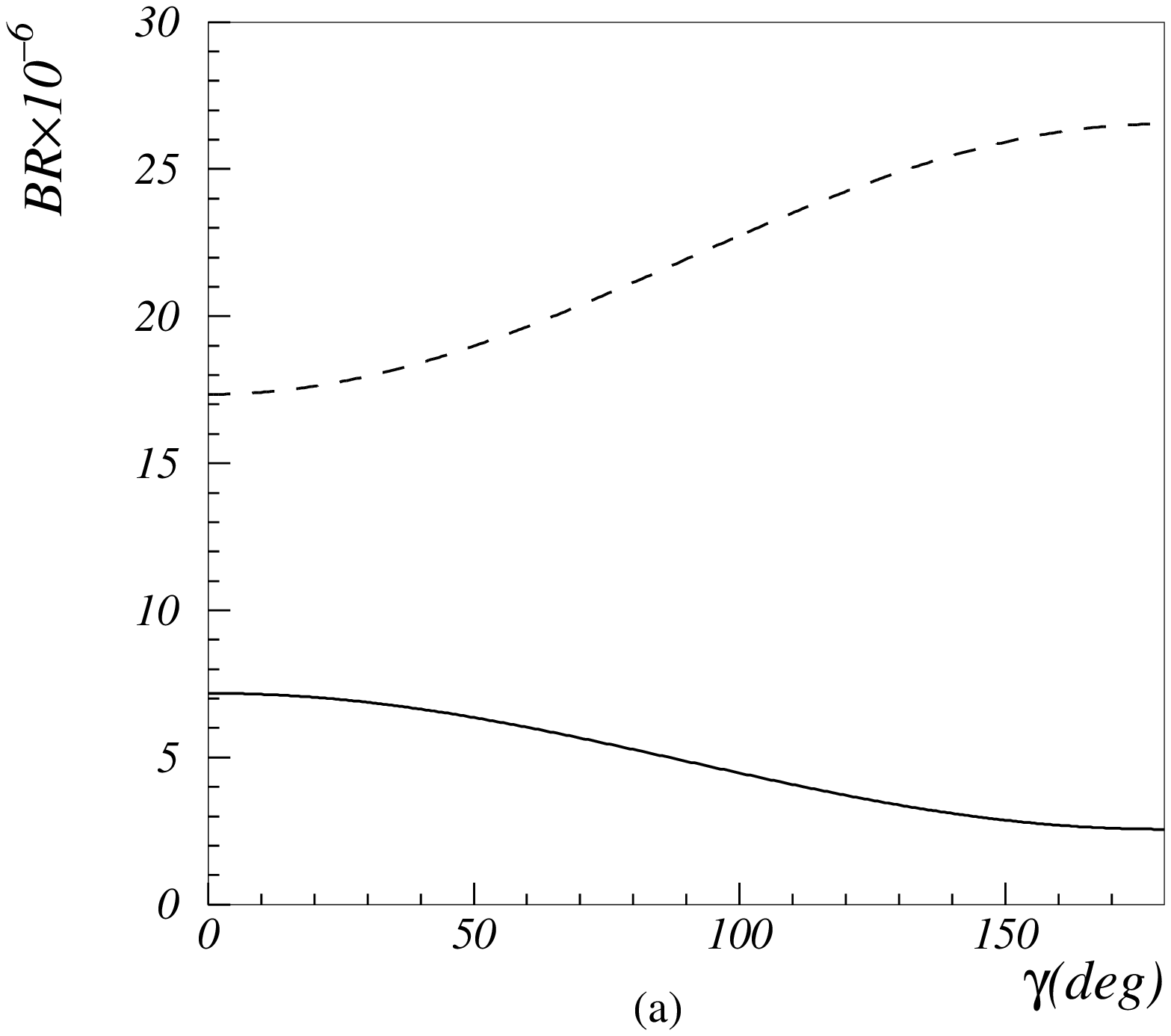,height=3.5in }
\end{figure}

\begin{figure}[tbp]
\hspace{3.5cm} \psfig{figure=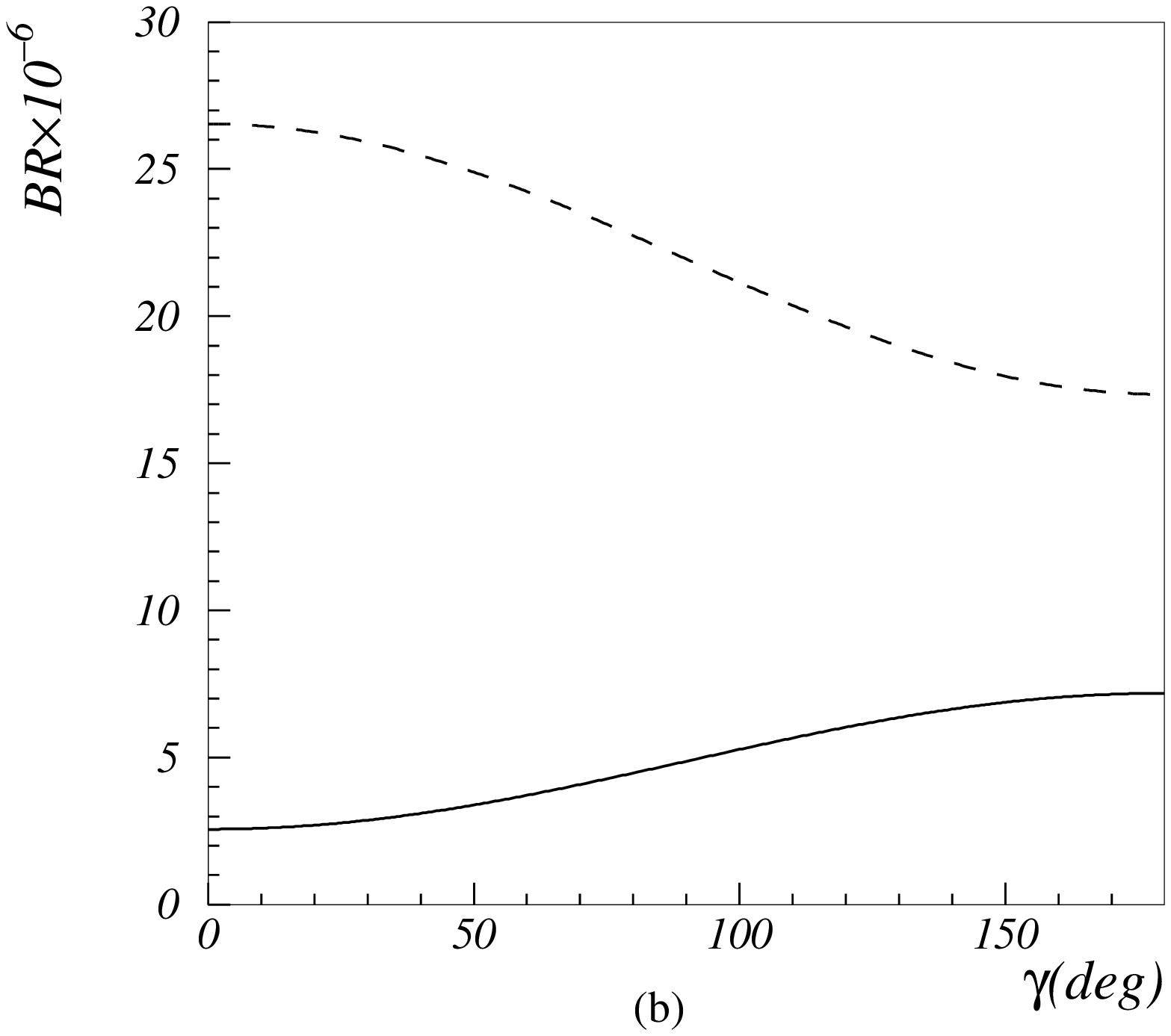,height=3.5in }
\caption{BRs (in units of $10^{-6}$) of $B_{s}\rightarrow K^+ K^-$ (dashed
line) and $B_{d}\rightarrow \pi^+ \pi^-$ (solid line) as a function of $%
\gamma$ with R(r)=8.0 and (a) $\delta=220^0$ and (b)
$\delta=40^0$. } \label{brpipi}
\end{figure}

\end{document}